\documentclass[twocolumn,A4]{article}
\usepackage[dvips]{graphics}
\begin{document}
\onecolumn
\begin{center}

{\bf{\Large Effect of isomers on quantum transport through molecular 
bridge system}}\\
~\\
Santanu K. Maiti$^{1,2,*}$ \\
~\\
{\em $^1$Theoretical Condensed Matter Physics Division, \\
Saha Institute of Nuclear Physics, \\
1/AF, Bidhannagar, Kolkata-700 064, India \\
$^2$Department of Physics, Narasinha Dutt College,
129, Belilious Road, Howrah-711 101, India} \\
~\\
{\bf Abstract}
\end{center}
Quantum transport for different models of isomer molecules attached
to two semi-infinite leads is studied on the basis of Green's function
technique. Electronic transport properties are significantly affected 
by (a) the relative position of the atoms in these molecules and (b) 
the molecule-to-lead coupling strength of these molecular bridge 
systems.
\vskip 1cm
\begin{flushleft}
{\bf PACS No.}: 73.23.-b; 73.63.-b; 81.07.Nb \\
~\\
{\bf Keywords}: Green's function; Isomer; Conductance; $I$-$V$ 
characteristic.
\end{flushleft}
\vskip 4.8in
\noindent
{\bf ~$^*$Corresponding Author}: Santanu K. Maiti

Electronic mail: santanu.maiti@saha.ac.in
\newpage
\twocolumn

\section{Introduction}

Molecular electronics and transport have attracted much more attention
since molecules constitute promising building blocks for future
generation of electronic devices. The transport through molecules was
first studied theoretically in $1974$~\cite{aviram}. Later several
numerous experiments~\cite{tali,metz,fish,reed1,reed2} have been 
performed through molecules placed between two electrodes with few 
nanometer separation. Full quantum mechanical treatment is required 
to characterize the transport through molecules. Some {\em ab initio} 
calculations~\cite{ven,cheng1,cheng2} are also used to study the 
current-voltage characteristics of a molecule. Electrical conduction 
through isomer molecules strongly depends on the relative position of 
the atoms in molecules and their coupling strength to the two leads, 
namely, left lead and right lead. Here we describe theoretically the 
electron transport in some specific models of isomer molecules. Based 
on the scanning probe technique measurement, conductance of molecular 
systems is directly possible~\cite{hong,tans,bock,don,cui,schon}. 
Theoretically there exist several formulations~\cite{baer1,baer2,baer3,
walc1,walc2,mol,roth} for the calculation of conductance based on 
Landauer formula and the seminal $1974$ paper of Aviram and
Ratner~\cite{aviram}. At much low temperatures and low bias voltages the
electron transport becomes coherent through the molecule. Here we
assume that the dissipation and equilibration processes occur only in the
two contacting leads and this approximation enables to describe the
propagation of an electron by means of single particle Green's function.
This theory is much more flexible than any other theoretical approach and
also applicable to any system described by a Hamiltonian with a localized
orbital basis. By using this method the electronic transport properties of
any system can be studied very easily with a very small computational 
cost. In that case we have to know only the Hamiltonian matrix for the 
molecule but no need to know anything about the electronic wave function.

Here, we investigate the conductance ($g$) and current-voltage ($I$-$V$)
characteristics of three different isomer molecules and compute the 
effect of these isomers on the electrical conduction.

The present article is organized as follows. In Section $2$, we describe 
the formulation of conductance $g$ by calculating the transmission 
probability $T$ and current $I$ for any finite size conducting system 
attached to two semi-infinite metallic leads by the use of Green's 
function technique. Section $3$ describes the conductance and 
current-voltage characteristics of single isomer molecules. 
Finally, we conclude our results in Section $4$.

\section{Formulation of $g$, $T$ and $I$: Green's function technique}

Here we give a brief description for the calculation of transmission
probability ($T$), conductance ($g$) and current ($I$) through a finite 
size conducting system attached to two semi-infinite metallic leads by 
the use of Green's function technique.

\begin{figure}[ht]
{\centering \resizebox*{8cm}{1.75cm}{\includegraphics{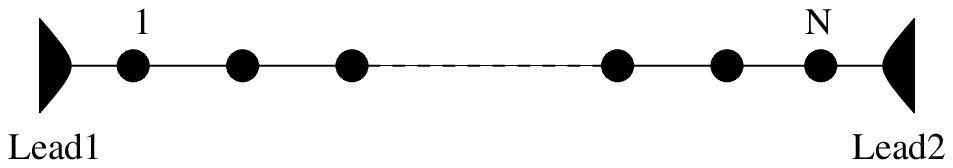}}\par}
\caption{Schematic view of a one-dimensional conductor with $N$ number of 
sites (filled circles) attached to two leads. The first and the last sites 
are labeled by $1$ and $N$, respectively.}
\label{dot}
\end{figure}

Let us first consider a one-dimensional conductor with $N$ number of atomic
sites (array of filled circles) connected to two semi-infinite leads, left 
lead and right lead as shown in Fig.~\ref{dot}. The conducting system in 
between the two leads
can be an array of some quantum dots, or a single molecule, or an array 
of some molecules, etc. At low voltages and low temperatures, the 
conductance of a conductor can be written through the Landauer 
conductance formula as~\cite{datta},
\begin{equation}
g=\frac{2e^2}{h}T
\label{land}
\end{equation}
where $g$ is the conductance and $T$ is the transmission probability of 
an electron through the conductor. This transmission probability can be 
expressed in terms of the Green's function of the conductor and the 
coupling of the conductor to the two leads in the following way,
\begin{equation}
T={\mbox {Tr}}\left[\Gamma_LG_c^r\Gamma_RG_c^a\right]
\label{trans1}
\end{equation}
where $G_c^r$ and $G_c^a$ are respectively the retarded and advanced Green's
function of the conductor. $\Gamma_L$ and $\Gamma_R$ are the coupling terms
between the conductor and the two leads. For the complete system, i.e., 
the system with the conductor and the two leads, the Green's function 
is defined as,
\begin{equation}
G=\left(\epsilon-H\right)^{-1}
\end{equation}
where $\epsilon=E+i\eta$. $E$ is the energy and $\eta$ is a very small 
number which can be put as zero in the limiting approximation. The above 
Green's function corresponds to the inversion of an infinite matrix which 
consists of the full system. It can be partitioned into different 
sub-matrices that correspond to individual sub-systems.

The Green's function for the conductor can be effectively expressed in 
the form,
\begin{equation}
G_c=\left(\epsilon-H_c-\Sigma_L-\Sigma_R\right)^{-1}
\label{grc}
\end{equation}
where $H_c$ is the Hamiltonian of the conductor sandwiched between the two
leads. The Hamiltonian of the conductor in the tight-binding framework
can be written within the non-interacting picture in this form,
\begin{equation}
H_c=\sum_i \epsilon_i c_i^{\dagger} c_i + \sum_{<ij>}t
\left(c_i^{\dagger}c_j + c_j^{\dagger}c_i \right)
\label{hamil1}
\end{equation}
where $c_i^{\dagger}$ ($c_i$) is the creation (annihilation) operator of an
electron at site $i$, $\epsilon_i$'s are the site energies and $t$ is the
nearest-neighbor hopping strength. In Eq.~\ref{grc},
$\Sigma_L=h_{Lc}^{\dagger} g_L h_{Lc}$ and $\Sigma_R=h_{Rc} g_R
h_{Rc}^{\dagger}$ are the self-energy terms due to the semi-infinite leads.
$g_L$ and $g_R$ are respectively the Green's function for the left and
right leads. $h_{Lc}$ and $h_{Rc}$ are the coupling matrices and they will
be non-zero only for the adjacent points in the conductor, $1$ and $N$
as shown in Fig.~\ref{dot}, and the leads respectively. The coupling terms
$\Gamma_L$ and $\Gamma_R$ for the conductor can be calculated through the
expression~\cite{datta},
\begin{equation}
\Gamma_{\{L,R\}}=i\left[\Sigma_{\{L,R\}}^r-\Sigma_{\{L,R\}}^a\right]
\end{equation}
where $\Gamma_{\{L,R\}}^r$ and $\Gamma_{\{L,R\}}^a$ are the retarded and
advanced self-energies respectively and they are conjugate with each
other. Datta {\em et al.}~\cite{tian} have shown that the self-energies
can be expressed in this form,
\begin{equation}
\Sigma_{\{L,R\}}^r=\Lambda_{\{L,R\}}-i \Delta_{\{L,R\}}
\end{equation}
where $\Lambda_{\{L,R\}}$ are the real parts of the self-energies for the
two leads. The imaginary parts $\Delta_{\{L,R\}}$ of the self-energies
represent the broadening of the energy level due to the coupling of the
conducting system with the two leads respectively. By calculating some
algebra these real and imaginary parts of the self-energies can also be 
determined in terms of coupling strength ($\tau_{\{L,R\}}$) between 
the conductor and the two leads, injecting energy ($E$) of the 
transmitting electrons and hopping strength ($t$) between nearest-neighbor 
sites in the leads. Thus the coupling terms $\Gamma_L$ and $\Gamma_R$ can 
be written in terms of the retarded self-energy as,
\begin{equation}
\Gamma_{\{L,R\}}=-2 Im\left[\Sigma_{\{L,R\}}^r\right]
\end{equation}
Thus by calculating the self-energies due to the left and right leads the
coupling terms $\Gamma_L$ and $\Gamma_R$ can be easily obtained and then
the transmission probability ($T$) will be calculated from the expression
as mentioned in Eq.~\ref{trans1}.

Since the coupling matrices $h_{Lc}$ and $h_{Rc}$ are non-zero only for
the adjacent points in the conductor, $1$ and $N$ as shown in Fig.~\ref{dot},
the transmission probability can be expressed in the following way,
\begin{equation}
T(E,V)=4\Delta_{11}^L(E,V) \Delta_{NN}^R(E,V)|G_{1N}(E,V)|^2
\label{trans2}
\end{equation}
For the sake of simplicity, here we assume that the entire voltage ($V$) 
is dropped across the conductor-electrode interfaces. 

The current through the bridge system can be written in the following 
form~\cite{datta},
\begin{equation}
I(V)=\frac{e}{\pi \hbar}\int \limits_{E_F-eV/2}^{E_F+eV/2} T(E,V) dE
\end{equation}
where $E_F$ is the Fermi energy of the conductor. Using the expression of 
$T(E,V)$ as in Eq.~\ref{trans2} the final form of $I(V)$ will be,
\begin{eqnarray}
I(V) &=& \frac{4e}{\pi \hbar}\int \limits_{E_F-eV/2}^{E_F+eV/2}
\Delta_{11}^L(E,V) \Delta_{NN}^R(E,V) \nonumber \\
& & \times ~|G_{1N}(E,V)|^2 dE
\label{curr}
\end{eqnarray}
Eq.~\ref{land}, Eq.~\ref{trans2} and Eq.~\ref{curr} are the final working
formulae for the calculation of conductance and current-voltage
characteristics respectively for any finite size conducting system 
connected to two semi-infinite leads.

Here, we shall describe conductance-energy and current-voltage 
characteristics by using the above formulations for some specific 
models of single isomer molecules. Throughout the paper, we use 
the units where $c=h=e=1$.

\section{Conductance and current-voltage characteristics of
isomer molecules}

\begin{figure}[ht]
{\centering \resizebox*{6cm}{12cm}{\includegraphics{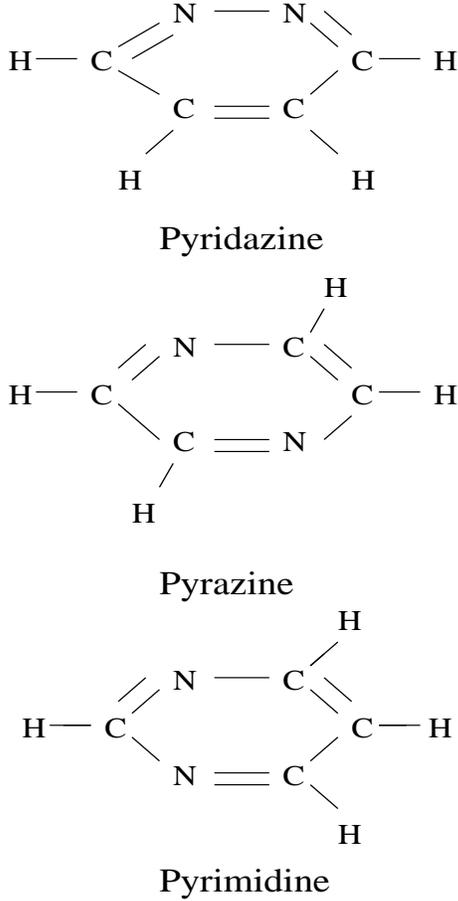}}\par}
\caption{Schematic representation of three different isomer molecules: 
pyridazine, pyrazine and pyrimidine those are attached to two leads.}
\label{isomer}
\end{figure}

Electron transport is strongly affected by the relative position of the 
atoms in single isomer molecules, and, to describe this effect we 
consider three different models of isomer molecule. Figure~\ref{isomer} 
corresponds to the schematic geometry of the three different isomer 
molecules. They are respectively defined as pyridazine, pyrazine and 
pyrimidine, where relative position of the two nitrogen atoms changes 
accordingly as shown in the figure. These single molecules are connected 
to two semi-infinite leads by thiol (S-H) groups (not shown here in the 
figure). In actual experimental setup, two leads made from gold (Au) 
are used and molecule attached to the leads by thiol (S-H) groups in 
the chemisorption technique where hydrogen (H) atoms remove and sulfur 
(S) atoms reside. Here we assume that the molecules are connected to 
the two leads via sulfur atoms by removing the extreme left and right 
hydrogen atoms of each molecule as shown in Fig.~\ref{isomer}. 

It should be noted that the electron transport is strongly affected by the
molecule to lead coupling strength and in this article we shall describe 
the isomeric effect on electron transport both for the weak and strong 
\begin{figure}[ht]
{\centering \resizebox*{8cm}{12.5cm}{\includegraphics{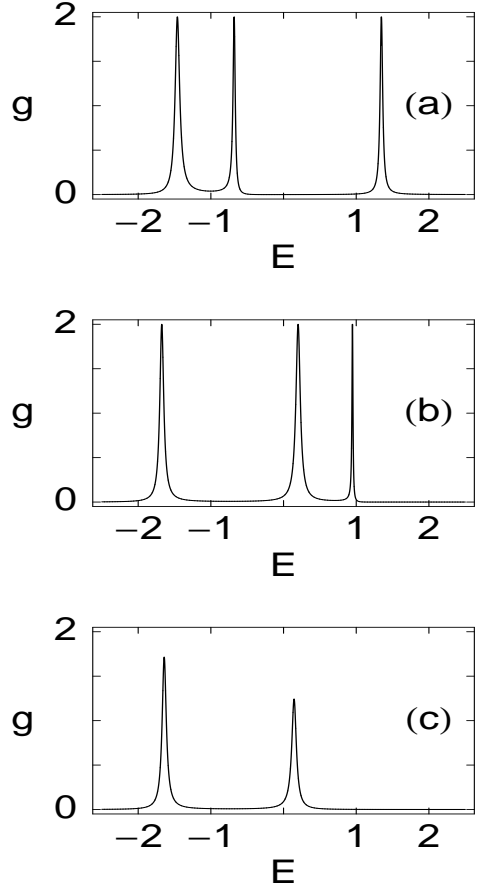}}\par}
\caption{Conductance $g$ as a function of injecting electron energy $E$ for
the weak coupling limit of three isomer molecules, where (a), (b) and (c) 
correspond to the pyridazine, pyrazine and pyrimidine molecules, 
respectively.}
\label{isocondlow}
\end{figure}
molecule-lead coupling limits. The weak coupling limit is mentioned through
the condition $\tau_{\{L,R\}}<<t$, while the strong coupling limit is 
denoted as $\tau_{\{L,R\}} \sim t$. $\tau_{\{L,R\}}$ is the molecule to lead 
coupling strength and $t$ is the hopping strength between nearest-neighbor
site in the molecule. Throughout the paper, we set the values of these 
parameters in the weak coupling limit as $\tau_L=\tau_R=0.5$, $t=3$, and, 
for the strong coupling limit these parameters are taken as 
$\tau_L=\tau_R=2$, $t=3$. Now, we try to characterize the 
conductance-energy and current-voltage characteristics of these 
single isomer molecules.

In Fig.~\ref{isocondlow}, we plot the conductance variation as a function of
energy $E$ of these three isomer molecules for the weak molecule to lead 
coupling limit. Figures~\ref{isocondlow}(a), (b) and (c) correspond to the 
conductance variation of pyridazine, pyrazine and pyrimidine molecules 
\begin{figure}[ht]
{\centering \resizebox*{8cm}{12cm}{\includegraphics{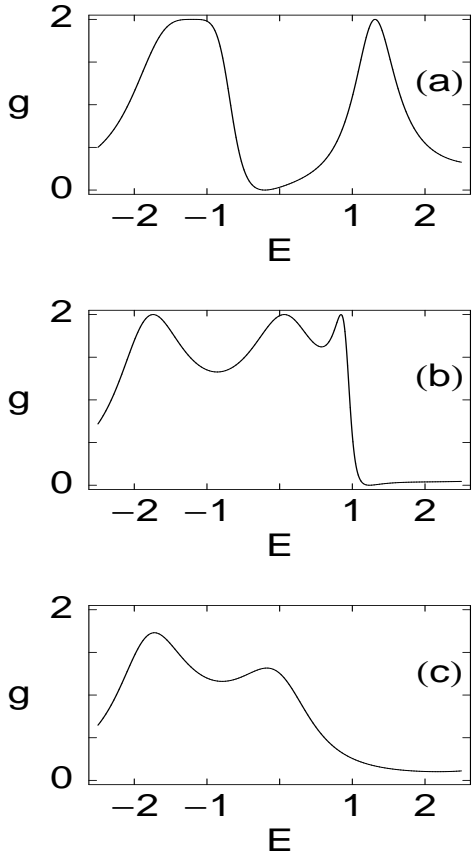}}\par}
\caption{Conductance $g$ as a function of injecting electron energy $E$ for
the strong coupling limit of three isomer molecules, where (a), (b) and (c) 
correspond to the pyridazine, pyrazine and pyrimidine molecules,
respectively.}
\label{isocondhigh}
\end{figure}
respectively. The conductance shows oscillatory behavior with sharp 
resonant peaks for some particular energy values, while for all other 
energy values it almost vanishes. This behavior can be understood as 
follows. Transmission of an electron through a molecule takes place 
only when the incident energy matches with anyone of the energy 
eigenvalues of the molecule. For this particular energy value the 
electron transmits almost ballistically from the source to drain, and 
accordingly, conductance shows a sharp peak. Now the electrons are carried 
from the left lead to right lead through isomer molecules and hence the 
electron waves propagating along the two arms
of the ring may suffer a phase shift between themselves, according to the
result of quantum interference between the various pathways that the electron
can take. Therefore, the probability amplitude of the electron across the
ring becomes strengthened or weakened. It emphasizes itself especially as 
transmittance cancellations, some peaks do not reach to unity anymore,
and anti-resonances in the transmission (conductance) spectrum. From 
Fig.~\ref{isocondlow}(a) and Fig.~\ref{isocondlow}(b) it is observed that at 
\begin{figure}[ht]
{\centering \resizebox*{7.5cm}{4.5cm}{\includegraphics{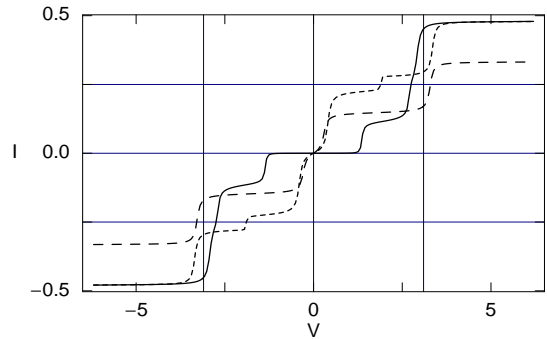}}\par}
\caption{Current $I$ as a function of the applied bias voltage $V$ for the
weak coupling limit of three isomer molecules, where the solid, dotted and 
dashed lines are respectively for the pyridazine, pyrazine and pyrimidine 
molecules.}
\label{isocurrlow}
\end{figure}
resonances the conductance shows the value $2$, i.e., for these resonances 
transmission probability $T$ goes to unity (since from Landauer formula we 
get $g=2T$ by putting $e=h=1$). On the other hand, Fig.~\ref{isocondlow}(c) 
shows that the conductance peak does not reach the value $2$ anymore. Though 
all these three molecules are attached 
symmetrically to the two leads, yet the probability amplitude across 
the pyrimidine molecule decreases compared to the other two molecules. So it 
can be emphasized that the electron transmission is strongly affected by the 
relative position of the two nitrogen atoms in these isomer molecules. Thus, 
different molecular structure can significantly affect the transport property 
even if they are isomer.

Another important feature observed from this Fig.~\ref{isocondlow} is that, 
for the pyrazine (Fig.~\ref{isocondlow}(b)) and pyrimidine 
(Fig.~\ref{isocondlow}(c)) molecules electron transmission starts at 
very low energy value, i.e., for low applied bias voltage $V$. On the other
hand, the pyridazine molecule starts electron transmission at quite higher 
energy value (see Fig.~\ref{isocondlow}(a)). Thus, the threshold bias voltage
for electron transmission also depends on the relative position of the two 
nitrogen atoms in these isomers.

To investigate the effect of conductance behavior in strong molecule to lead 
coupling limit, we plot the results of $g$ as a function of energy $E$ for 
these three isomer molecules in Fig.~\ref{isocondhigh}, where (a), (b) and
(c) are respectively for the pyridazine, pyrazine and pyrimidine molecules.
From these curves it is observed that the resonant peaks get substantial
widths and electron transmission takes place for wide range of energy values
through these molecular bridge systems. This is due to the substantial 
broadening of the molecular energy levels caused by the strong coupling of
the molecule to the two semi-infinite leads~\cite{datta}. These effects 
have strong dependence on the current-voltage characteristics which we 
are going to discuss in the following parts. 

Now, we study the current-voltage characteristics of these molecular bridge 
systems. The current $I$ is computed by the integration procedure of the 
\begin{figure}[ht]
{\centering \resizebox*{7.5cm}{4.5cm}{\includegraphics{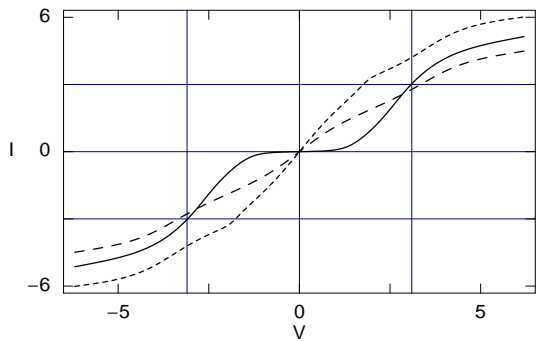}}\par}
\caption{Current $I$ as a function of the applied bias voltage $V$ for the
strong coupling limit of three isomer molecules, where the solid, dotted and 
dashed lines are respectively for the pyridazine, pyrazine and pyrimidine 
molecules.}
\label{isocurrhigh}
\end{figure}
transmission function $T$. The behavior of the transmission function $T$
with energy $E$ is exactly similar (differ only in magnitude by the factor 
$2$ due to the existence of the relation $g=2T$) to that as presented in 
Fig.~\ref{isocondlow} and Fig.~\ref{isocondhigh}. Figure~\ref{isocurrlow} 
represents some current-voltage characteristics of the three molecules 
in the weak molecular coupling limit, where the solid, dotted and dashed 
lines give the current for the pyridazine, pyrazine and pyrimidine 
molecules, respectively. The current shows staircase-like behavior with 
sharp steps. This is due to the discreteness of the molecular resonances 
as shown in Fig.~\ref{isocondlow}. From this figure it is clearly observed 
that, both the pyrazine and pyrimidine molecules show finite non-zero 
value of current (dotted and dashed curves) for very small applied bias 
voltages, while the pyridazine molecules gives non-zero value of current 
(solid curve) at higher bias voltages.

In the presence of strong molecule to lead coupling limit, the 
staircase-like behavior disappears and current varies quite continuously 
with the applied bias voltage. The variation of the current for these 
three isomer molecules in the strong coupling limit is plotted in 
Fig.~\ref{isocurrhigh}, where the solid, dotted and dashed lines 
correspond to the same meaning as earlier. For this coupling limit, 
molecular resonances get substantial width (Fig.~\ref{isocondhigh}) 
and since the current is computed by the integration procedure of 
the transmission function $T$, it gets a continuous variation with 
the applied bias voltage $V$. Similar to the weak coupling case, the 
pyrimidine molecule gives non-zero value of the current at sufficient 
higher bias voltages compared to the other two molecules in this strong 
coupling case. Another important observation is that, in the strong 
coupling limit current amplitude becomes very large than that of the 
current amplitude in the weak coupling limit for these molecular 
bridge systems. 

\section{Concluding remarks}

To conclude, we have investigated the effect of isomers and molecule to
lead coupling strength on electron transport through some models of 
different types of isomer molecules. Depending on the relative position 
of the two nitrogen atoms in these isomer molecules, threshold bias 
voltage for electronic conduction changes which provides an important 
signature for the fabrication of different molecular bridge systems. 
The conductance shows fine resonant peaks which gives the staircase-like 
behavior with sharp steps in current for the weak coupling limit, while 
in the strong coupling limit, the resonant peaks get substantial widths, 
and accordingly, the current varies almost continuously with the applied 
bias voltage.   

\vskip 0.3in
\noindent
{\bf\Large Acknowledgments}
\vskip 0.2in
\noindent
It is my pleasure to thank Atikur Rahman and Prof. S. N. Karmakar for 
many helpful comments and suggestions.


\begin{thebibliography}{99}

\bibitem{aviram} A. Aviram and M. Ratner, Chem. Phys. Lett. \textbf{29}
(1974) 277.
\bibitem{ven} M. Di Ventra, S. T. Pantelides, and N. D. Lang, Phys. Rev.
Lett. \textbf{84} (2000) 979.
\bibitem{cheng1} W. W. Cheng, H. Chen, R. Note, H. Mizuseki, and Y. Kawazoe,
Physica E \textbf{25} (2005) 643.
\bibitem{cheng2} W. W. Cheng, Y. X. Liao, H. Chen, R. Note, H. Mizuseki,
and Y. Kawazoe, Phys. Lett. A \textbf{326} (2004) 412.
\bibitem{tali} T. Dadosh, Y. Gordin, R. Krahne, I. Khivrich, D. Mahalu,
V. Frydman, J. Sperling, A. Yacoby, and I. Bar-Joseph, Nature \textbf{436}
(2005) 677.
\bibitem{metz} R. M. Metzger {\em et al.}, J. Am. Chem. Soc. \textbf{119}
(1997) 10455.
\bibitem{fish} C. M. Fischer, M. Burghard, S. Roth, and K. V. Klitzing,
Appl. Phys. Lett. \textbf{66} (1995) 3331.
\bibitem{reed1} J. Chen, M. A. Reed, A. M. Rawlett, and J. M. Tour, Science
\textbf{286} (1999) 1550.
\bibitem{reed2} M. A. Reed, C. Zhou, C. J. Muller, T. P. Burgin, and J. M.
Tour, Science \textbf{278} (1997) 252.
\bibitem{hong} S. Hong, W. Tian, J. Henderson, S. Datta, C. P. Kubiak, and
R. Reifenberger, Superlat. Microstruct. \textbf{28} (2000) 289.
\bibitem{tans} S. J. Tans, R. M. Verschueren, and C. Dekker, Nature
\textbf{393} (1998) 49.
\bibitem{bock} M. Bockrath, D. H. Cobden, P. L. McEuen, N. G. Chopra,
A. Zettl, A. Thess, and R. E. Smalley, Science \textbf{275} (1997) 1922.
\bibitem{don} Z. J. Donhauser {\em et. al.}, Science \textbf{292}
(2001) 2303.
\bibitem{cui} X. D. Cui, A. Primak, X. Zarate {\em et. al.}, Science
\textbf{294} (2001) 571.
\bibitem{schon} J. H. Schon, H. Meng, and Z. N. Bao, Science \textbf{294}
(2001) 2138.
\bibitem{baer1} R. Baer and D. Neuhauser, Chem. Phys. \textbf{281} (2002) 
353.
\bibitem{baer2} R. Baer and D. Neuhauser, J. Am. Chem. Soc. \textbf{124}
(2002) 4200.
\bibitem{baer3} D. Walter, D. Neuhauser, and R. Baer, Chem. Phys.
\textbf{299} (2004) 139.
\bibitem{walc1} K. Walczak, Cent. Eur. J. Chem. \textbf{2} (2004) 524.
\bibitem{walc2} K. Walczak, Phys. Stat. Sol. (b) \textbf{241} (2004) 2555. 
\bibitem{mol} {\em Molecular Electronics-Science and Technology}, edited by
A. Aviram and M. Ratner (N. Y. Acad. Sci., N. Y., 1998) (volume \textbf{852}
of Ann. N. Y. Acad. Sci.).
\bibitem{roth} S. Roth and C. Joachim, {\em Atomic and Molecular Wires}
(Kluwer, Dordrecht, 1997).
\bibitem{datta} S. Datta, {\em Electronic transport in mesoscopic systems},
Cambridge University Press, Cambridge (1997).
\bibitem{tian} W. Tian, S. Datta, S. Hong, R. Reifenberger, J. I. Henderson,
and C. I. Kubiak, J. Chem. Phys. \textbf{109} (1998) 2874.

\end{thebibliography}
\end{document}